\documentclass[reprint, aps, prl, floatfix,superscriptaddress]{revtex4-1}
\usepackage{graphicx}
\usepackage{dcolumn}
\usepackage{bm}
\usepackage{amssymb}
\usepackage{amsmath}
\usepackage{amsthm}
\usepackage{mathrsfs}
\usepackage{amssymb}
\usepackage{amsfonts}
\usepackage{commath}
\usepackage{subfigure}
\usepackage{braket}
\usepackage{natbib}
\usepackage{float}
\usepackage{color}
\usepackage{siunitx}
\usepackage[colorlinks=true, allcolors=blue]{hyperref}
\usepackage{hyphenat}
\usepackage{footnote}
\usepackage{notes2bib}
\usepackage{xcolor}
\usepackage{threeparttable}
\usepackage{lineno}

\newcommand{\beq}{\begin{equation}}
\newcommand{\eeq}{\end{equation}}

\begin{document}

\title{Universal phase correction for quantum state transfer in \\one-dimensional topological spin chains}

\author{Tian Tian}
\email{phystian@sxu.edu.cn}
\affiliation{State Key Laboratory of Quantum Optics Technologies and Devices, Institute of Opto-Electronics, Shanxi University, Taiyuan 030006, China}
\affiliation{Collaborative Innovation Center of Extreme Optics, Shanxi University, Taiyuan 030006, China}

\begin{abstract}
Gap-protected topological channels are a promising way to realize robust and high-fidelity state transfer in quantum networks. Although various topological transfer protocols based on the Su-Schrieffer-Heeger model or its variants have been proposed, the phase accumulation during the evolution, as an essential aspect, is underestimated. Here, by numerically studying the phase information of quantum state transfer (QST) in one-dimensional (1D) topological spin chains, we uncover a universal phase correction $\phi_0 =(N-1)\pi/2$ for both adiabatic and diabatic topological schemes. Interestingly, the site-number-dependent phase correction satisfies $\mathbb{Z}_{4}$ symmetry and is equally effective for perfect mirror transmission in spin chains. Our work reveals a universal phase correction in 1D topologically protected QST, which will prompt a reevaluation of the topological protection mechanism in quantum systems.
\end{abstract}

\maketitle

\textit{Introduction.}---
Realization of high-fidelity quantum state transfer (QST) from a sender to a receiver node is an essential requirement for quantum communication and quantum computation~\cite{Nielsen2012}.
For short distance QST, how to construct transfer channels through a network of qubits in a quantum processor is a critical problem.
Originating from an unmodulated spin chain~\cite{Bose2003,Bose2007}, various QST schemes have been reported in the last two decades~\cite{DiFranco2008,Chudzicki2010,PembertonRoss2011,Eldredge2017}. 
Among these, the notable protocols of perfect state transfer~\cite{Cook1979,Christandl2004,Plenio2004} in multi-qubit networks have been demonstrated in quantum systems including nuclear spins~\cite{Zhang2005}, photonic waveguides~\cite{Bellec2012,PerezLeija2013,Chapman2016}, nanomechanical oscillators~\cite{Tian2020} and superconducting qubits~\cite{Li2018}.
However, these transfer schemes depend on precise individual couplings and accurate timing of dynamical evolution. Practical imperfections in quantum networks and timing errors in manipulation degrade the transfer fidelity. This motivates the need to realize robust high-fidelity QST.

Due to the disorder-immunity property of edge state transport, a series of gap-protected topological transfer mechanisms has been proposed in recent years~\cite{Yao2013,Almeida2016,Dlaska2017,Lang2017,Mei2018,Longhi2019a,
Longhi2019,Palaiodimopoulos2021,Yuan2021,Tian2024,Guo2024,Han2024}. These mechanisms require qubits to be coupled in specific topological configurations that form chain-like structures, along with temporal modulation to achieve robust high-fidelity state transfer. Crucially, the temporal evolution of any quantum state under a Hamiltonian inevitably induces phase accumulation, which subsequently compromises transfer fidelity. However, discussions of accumulated phases remain notably absent in current protocols for topologically protected QST.

Here, we first revisit QST in spin chains, elucidating the critical role of the phase of transition amplitude and the equivalence within the single-excitation subspace. We then numerically investigate transport processes in various topological transport protocols, focusing on phase and magnitude under disorder perturbations.
Our findings reveal that upon completion of a full transfer period, a universal phase factor accumulates at the receiver, which depends solely on spin numbers and exhibits $\mathbb{Z}_{4}$ symmetry.
Remarkably, this phase relationship remains valid even for non-adiabatic topological transport protocols.
Finally, we demonstrate that disorder perturbations in the Rice-Mele model-based transfer scheme induce chaotic phase factors, rendering these approaches more suitable for classical signal transmission than for quantum state transfer.

\begin{figure*}[t]
\centering
\includegraphics[width=1.8\columnwidth]{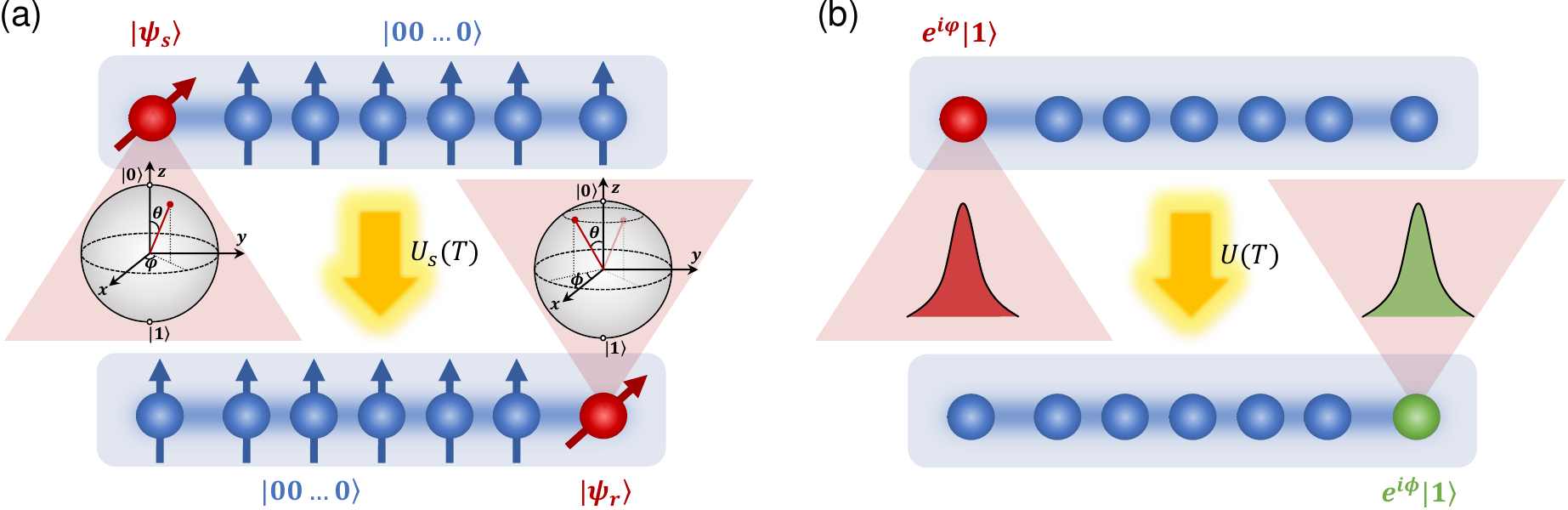}
\caption{Quantum state transfer in spin chains. (a) The initial pure state $\lvert \psi_s\rangle=\cos\frac{\theta}{2}\lvert 0\rangle+e^{i\varphi}\sin\frac{\theta}{2} \lvert 1\rangle$ at the sender is transferred to the receiver after an evolutionary period $T$ in a spin chain. Due to the phase accumulation during the evolution $U_{s}(t)=e^{-iHt}$, the final state at the receiver $\lvert \psi_r\rangle=\cos\frac{\theta}{2}\lvert 0\rangle+e^{i\phi}\sin\frac{\theta}{2} \lvert 1\rangle$ can be reproduce the initial state $\lvert \psi_s\rangle$ after operating a phase correction $e^{i(\varphi-\phi)}$ at the excited part. The initial and final quantum states are represented by red points on the Bloch sphere. (b) In the single-excitation subspace, the process of quantum state transfer in (a) is reduced to an initial excitation transmission in a bosonic (spinless) chain. The phase information can be obtained by the transition amplitude $A(T)=\langle N \lvert U(T)\lvert 1 \rangle$, where $U(T)=\hat{\mathcal{T}}\exp [-i\int_{0}^{T}d\tau \hat{H}(\tau)]$ is the time-evolution operator.
}
\label{Fig1}
\end{figure*}

\textit{Phase accumulation of QST.}---
We consider the transfer of a quantum state $|\psi\rangle_{s}=\cos\frac{\theta}{2}|0\rangle+e^{i\varphi}\sin\frac{\theta}{2}|1\rangle$ from the sender to the receiver qubit in a spin chain. As shown in Fig.~\ref{Fig1}(a), the initial state of this spin system is $|\Psi(0)\rangle=\cos\frac{\theta}{2}|\textbf{0}\rangle+e^{i\varphi}\sin\frac{\theta}{2}|\textbf{1}\rangle$,
where $|\textbf{0}\rangle=|000\ldots 0 \rangle$ corresponds to the ground state with an eigenenergy of $E_0=0$ and $|\textbf{1}\rangle=|100\ldots 0 \rangle$ corresponds to the first spin points upward while all other spins remain downward. After a specific evolution time $t$ governed by the Hamiltonian 
\begin{equation}
H=\sum_n J_{n}[\sigma_{n}^{x}\sigma_{n+1}^{x}+\sigma_{n}^{y}\sigma_{n+1}^{y}],
\label{Hamiltonian1}
\end{equation}
the quantum state of the $N$-th (receiver) spin is~\cite{Bose2003} 
\begin{equation}
|\psi_{r}(t)\rangle=\cos\frac{\theta}{2}|0\rangle+A(t)e^{i\varphi}\sin\frac{\theta}{2}|1\rangle,
\end{equation}
where $\sigma^{x,y}$ are the Pauli matrices and $A(t)=\langle \textbf{N} \lvert e^{-iHt} \lvert \textbf{1} \rangle$ is the transition amplitude for an excitation from the sender to the receiver.

Assuming that after time $t=T$, the quantum state at the receiver $|\psi_{r}(T)\rangle$ approaches the input state $|\psi_{s}\rangle$. Considering all pure states in the Bloch sphere as initial states, one obtains the average transfer fidelity~\cite{Bose2007,Bose2003,Petrosyan2010}
\begin{equation}
\mathcal{F}=\frac{1}{2}+\frac{|A(T)|^2}{6}+\frac{|A(T)|\cos{\gamma}}{3},
\label{Fidelity}
\end{equation}
where $\gamma=\arg \{A(T)\}$. It is evident that achieving high transfer fidelity requires not only the magnitude of the transition amplitude to approach 1 but also the phase of the transition amplitude to be either integer multiples of $2\pi$ or a predetermined fixed value~\cite{Bose2003,Yung2006,Petrosyan2010}. More importantly, since this phase is induced by single-excitation dynamics, we can thereby redefine the transition amplitude $A(T)=\langle N \lvert U(T)\lvert 1 \rangle$ within the single-excitation subspace, where $\lvert j \rangle$ stands for the excitation at $j$-th site, as shown in Fig.~\ref{Fig1}(b).

Therefore, when implementing topological transport protocols for QST in spin chains, the phase accumulation of the transition amplitude $\gamma=\arg\{A\}$ must be a predetermined fixed value. Otherwise, even if $|A|=1$, the average fidelity would be limited to the classical threshold of $\mathcal{F}=2/3$ due to the random phase $\gamma$. However, this critical requirement remains underexplored in existing protocols, especially under disorder-induced perturbations.

\begin{figure}[t]
\centering
\includegraphics[width=1.0\columnwidth]{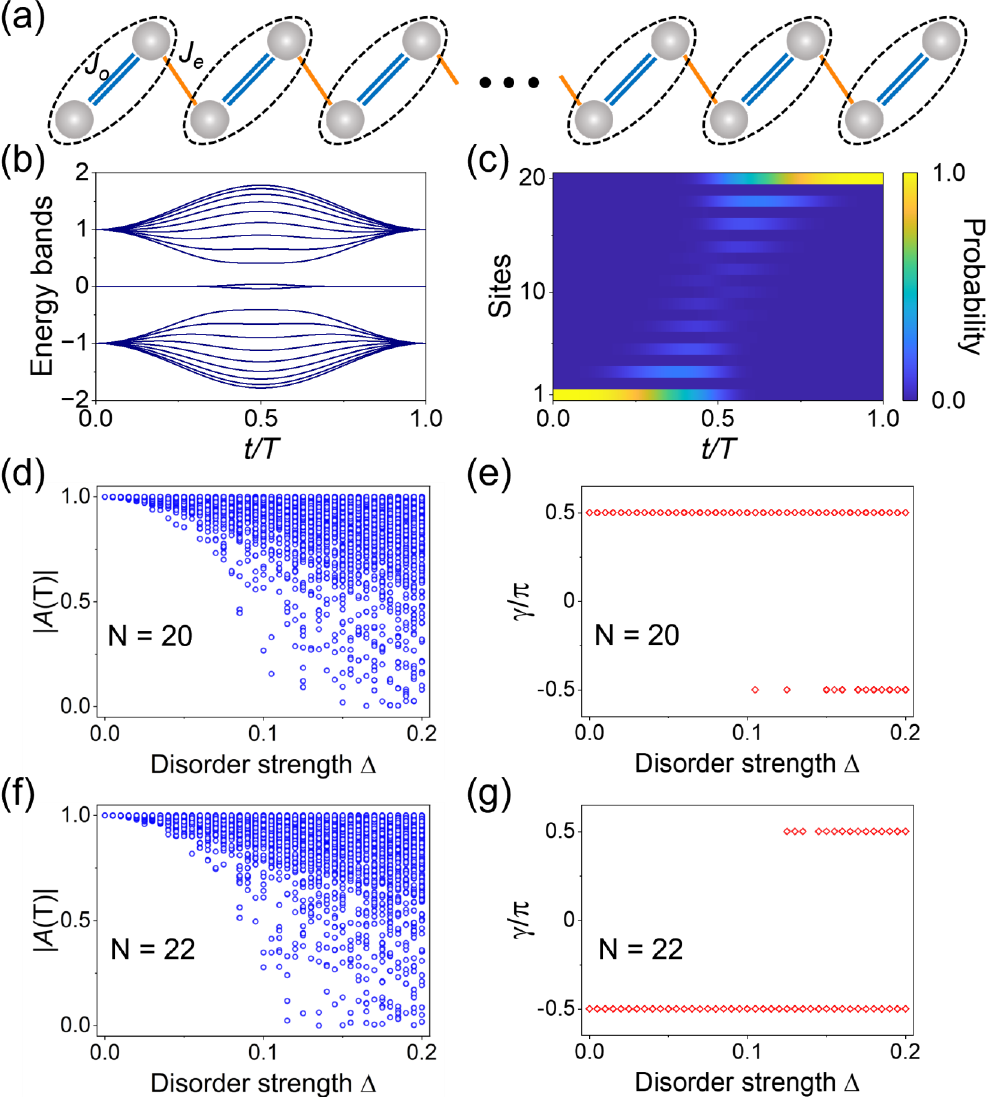}
\caption{Normal SSH transport. (a) An SSH chain with even sites.  The  intra-cell hopping $J_{\rm o}(t)=(1-\epsilon)\sin^{2}(\pi t/T)$ while  inter-cell hopping maintains $J_{\rm e}(t)=1$. (b,c) Energy bands and transfer probability of the dynamical evolution for a SSH chain with $N=20$ and $\epsilon=0.2$. (d-g) The the magnitude $|A|$ and the accumulated phase $\gamma=\arg \{A\}$ of the transition amplitude at time $t=T$ in the presence of random disorder $\delta J_n $. At each disorder strength, we numerically perform 200 independent disorder realizations. We set $T=200$ for $N=20$ and $T=260$ for $N=22$.}
\label{Fig2}
\end{figure}

\begin{figure*}[t]
\centering
\includegraphics[width=2.0\columnwidth]{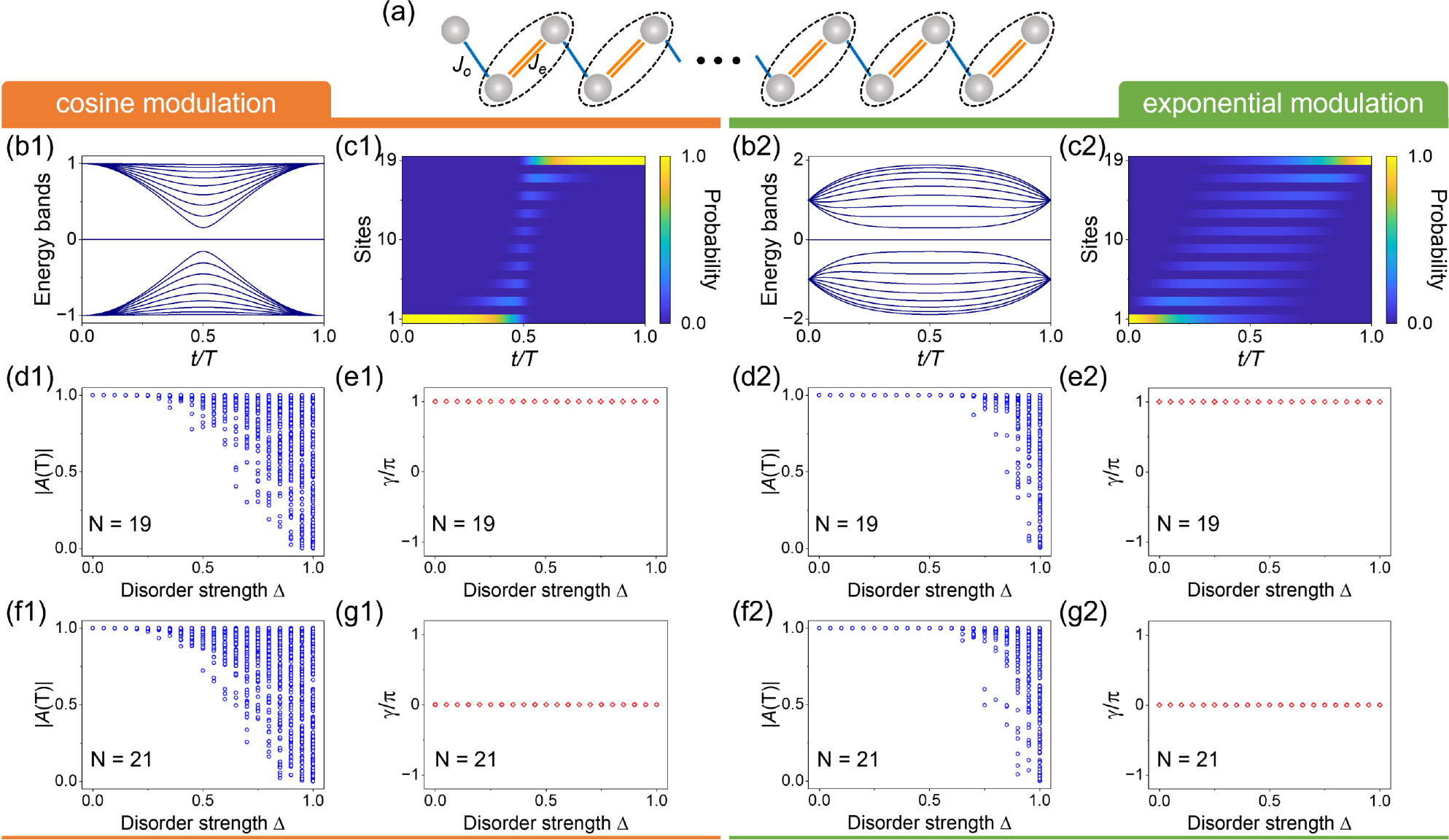}
\caption{Edge-defect topological transport. (a) An SSH chain with an edge defect.  (b1,c1) Energy bands and transfer probability of the edge defect SSH model with cosine modulations  $2J_{\rm o}(t)=1-\cos(\pi t/T)$, $2J_{\rm e}(t)=1+\cos(\pi t/T)$.  (b2,c2) Energy bands and transfer probability of the edge defect SSH model with exponential modulations $J_{\rm o}(t)=(1-e^{-\alpha t/T})/(1-e^{-\alpha})$ and $J_{\rm e}(t)=[1-e^{-\alpha (1-t/T)}]/(1-e^{-\alpha})$. The magnitude (d1,d2,f1,f2) and the accumulated phase (e1,e2,g1,g2) of the transition amplitude at $t=T$  in the presence of random disorder $\delta J_n $.  At each disorder strength, we numerically generate 200 independent disorder realizations. In above calculations, the total evolution time $T=1000$ and the parameter $\alpha=6$.}
\label{Fig3}
\end{figure*}

\begin{figure}[t]
\centering
\includegraphics[width=1.0\columnwidth]{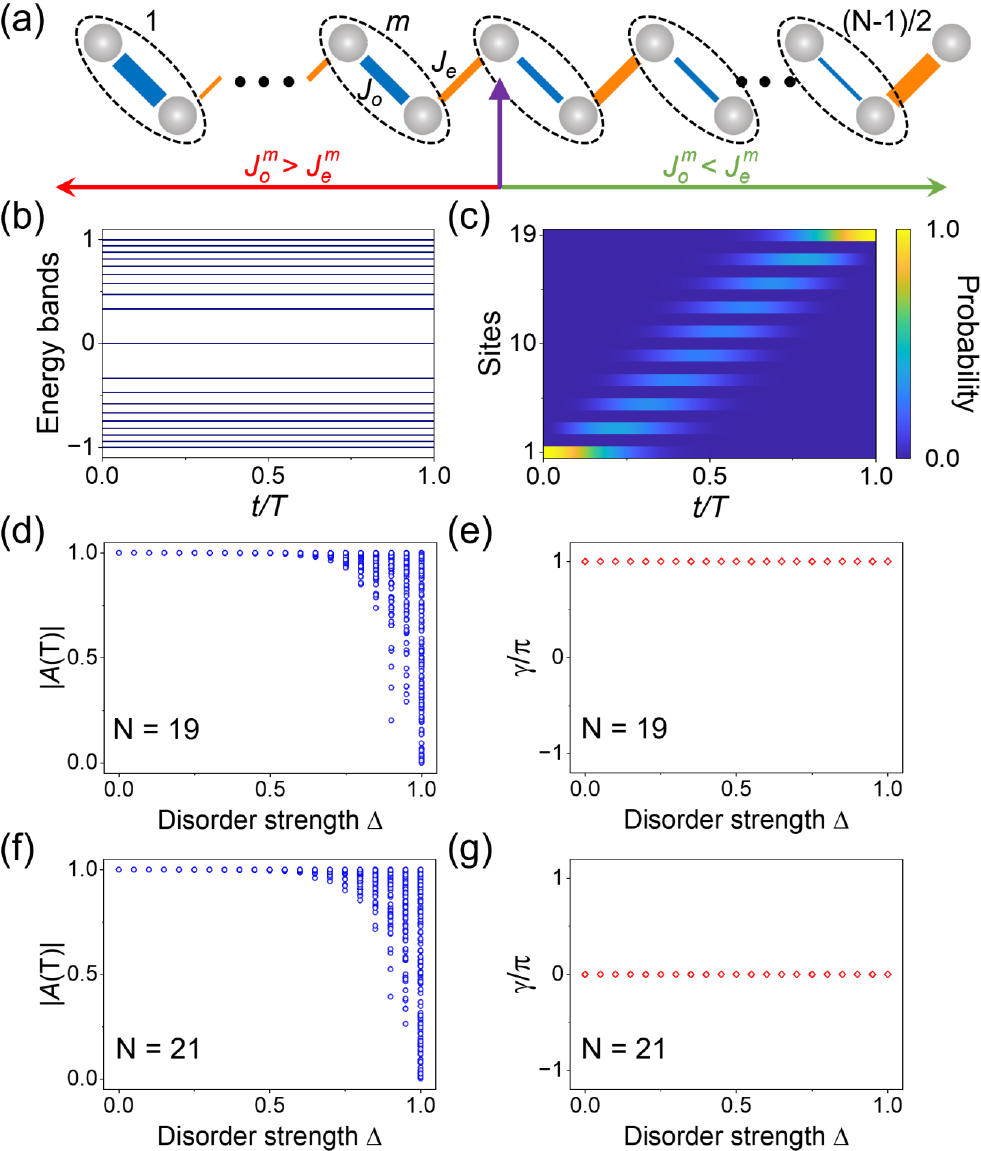}
\caption{Topological interface transport with square-root interactions. (a) The interface model includes two distinct topological phases. The SSH chain before the interface (purple arrow) is topologically trivial since $J_{\rm o}^{m}>J_{\rm e}^m{}$ for any unit cell.  After the interface, the SSH chain is non-trivial because $J_{\rm o}^{m}<J_{\rm e}^{m}$. Here, $m$ is the index of unit cells and $N_{c}$ is the total cells.
The inhomogeneous hopping terms $J_{\rm o}^{m}(t)=\sin(\pi t/2T)\sqrt{(N_{c}-m+1)/N_{c}}$ and $J_{\rm e}^{m}(t)=\cos(\pi t/2T)\sqrt{m/N_{c}}$. (b,c) Energy bands and the dynamical evolution of the topological interface transfer with $N=19$. (d-g) The magnitude and the phase of the transition amplitude of the receiver site at the moment $t=T$ in the presence of random disorder $\delta J_n $.  For each disorder strength, we generate 200 independent disorder realizations. In above calculations, we set total evolution time $T=1000$.}
\label{Fig4}
\end{figure}

\begin{figure}[t]
\centering
\includegraphics[width=1.0\columnwidth]{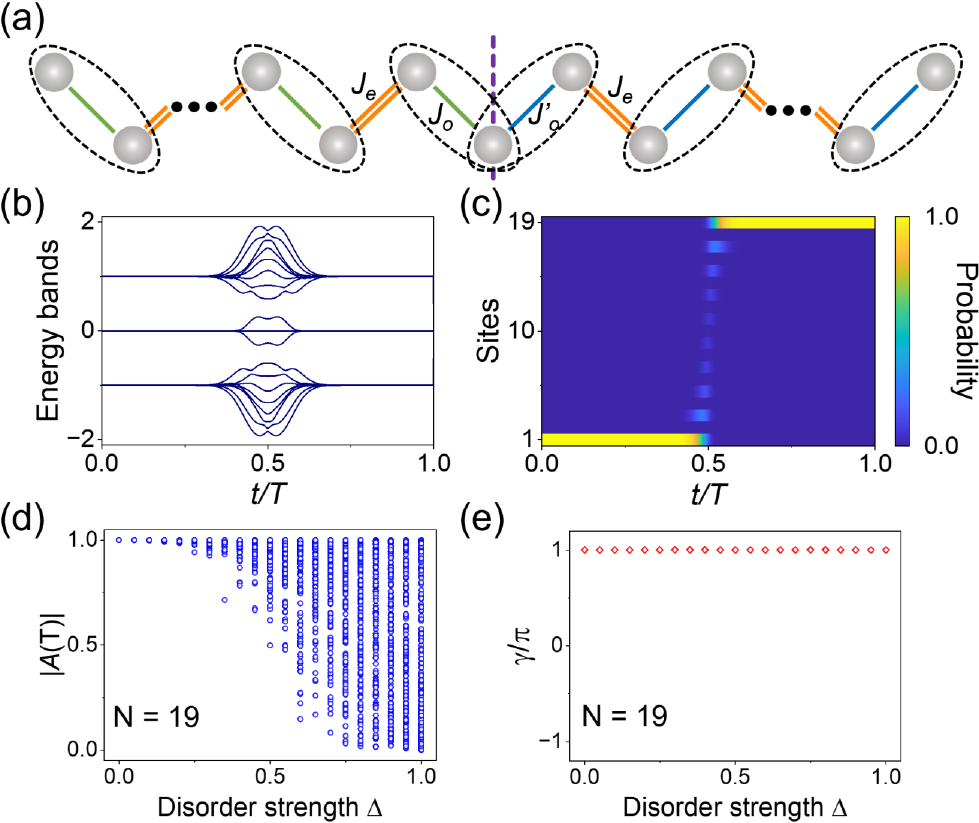}
\caption{Topological interface transport with Gaussian modulations. (a) The topological interface model includes two normal SSH chains. The number of total sites $N=2M-1$ is odd because each single SSH lattice comprises even sites ($M$ is even). Two intra-cell hopping terms are $J_{\rm o}(t)=\exp [-(t-\delta/2-T/2)^2/w^2)]$ and $J_{\rm o}^{'}(t)=\exp [-(t+\delta/2-T/2)^2/w^2)]$, respectively. At the same time, the inter-cell hopping terms remain constant $J_{\rm e}(t)=1$. (b,c) Energy bands and transfer probability of the Gaussian-modulated topological interface transport. The time delay $\delta=50$, Gaussian width $w=70$ and total transfer period $T=1000$ for all simulations. (d,e) The magnitude and phase of the transition amplitude $A(T)$ at the receiver site in the presence of random disorder $\delta J_n $.  For each disorder strength, we generate 200 independent disorder realizations.}
\label{Fig5}
\end{figure}

\begin{figure}[t]
\centering
\includegraphics[width=1.0\columnwidth]{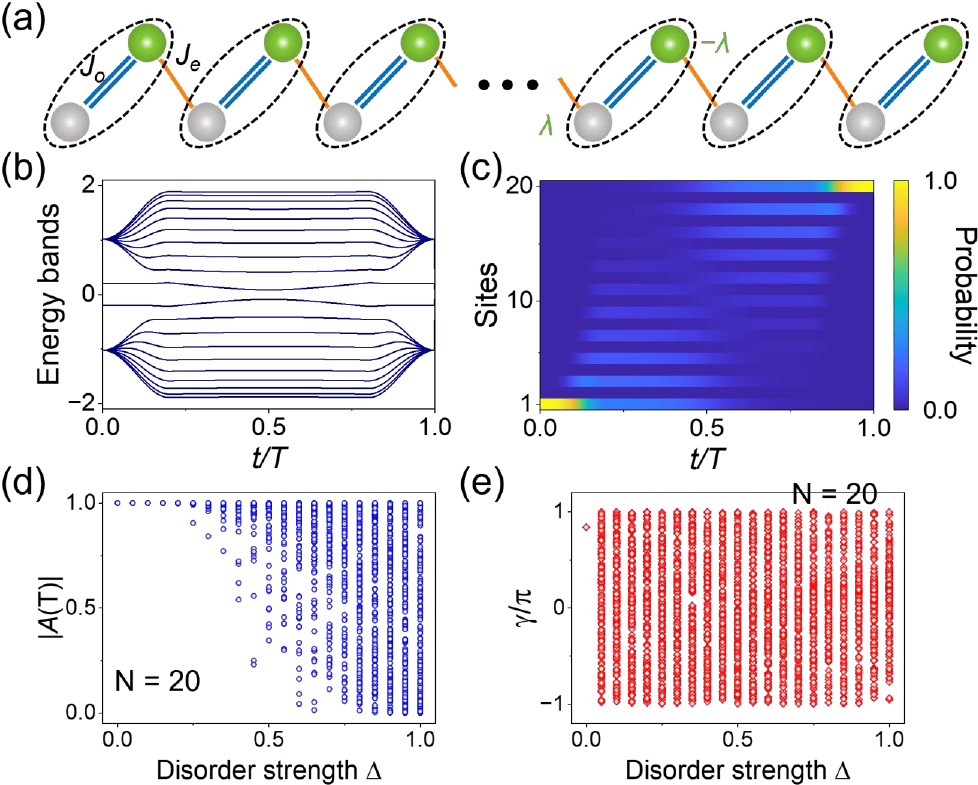}
\caption{Rice-Mele topological transport. (a) A standard SSH chain with staggered on-site terms. The time-varying intra-cell hopping $J_{\rm o}$ and the on-site fields $\lambda(t)$ follow the three-stage modulation in Equations~(\ref{RM-Jo}, \ref{RM-lambda}) while the inter-cell hopping keeps constant $J_{\rm e}(t)=1$. (b,c) Energy bands and transfer probability of the Rice-Mele scheme with $N=20$. (d,e) The magnitude and phase of the transition amplitude $A(T)$ in the presence of random disorder $\delta J_n $.  For each disorder strength, we generate 200 independent disorder realizations. We set the total evolution time $T=1000$ in all calculations. }
\label{Fig6}
\end{figure}

\begin{figure}[t]
\centering
\includegraphics[width=1.0\columnwidth]{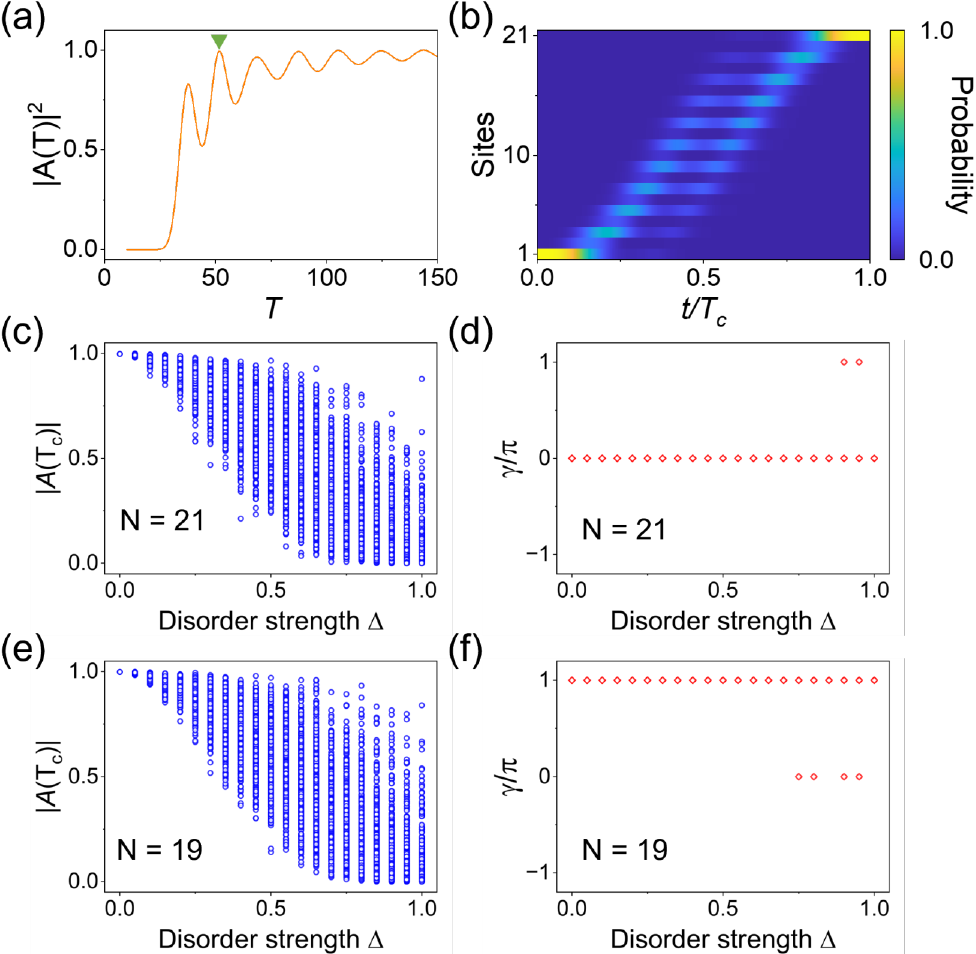}
\caption{Nonadiabatic topological transfer. (a) The transfer probability of the topological interface transport with square-root interactions under different transfer periods. (b) The excitation evolution process with $N=21$ for a nonadiabatic period indicated by the green triangle in (a). The magnitude (c,e) and the phase (d,f) of the nonadiabatic transition amplitude $A(T_{c})$ at the receiver site in the presence of random disorder $\delta J_n$. $T_{c}=52$ for $N=21$ and $T_{c}=48$ for $N=19$. For each disorder strength, we randomly generate 500 independent disorder realizations.}
\label{Fig7}
\end{figure}

\textit{Normal SSH transport.}---
As shown in Fig.~\ref{Fig2}(a), we consider an adiabatic topological  scheme based on the standard Su-Schrieffer-Heeger (SSH) model~\cite{Su1979}. Its Hamiltonian reads 
\begin{equation}
H(t)=J_{n}(t)\sum_{n}(a_n^{\dag}a_{n+1}+a_{n+1}^{\dag}a_n).
\label{H_even}
\end{equation}
Notably, this chain contains even sites and two alternating hopping terms $J_{n}=J_{\rm o}(t)=(1-\epsilon)\sin^{2}(\pi t/T)$ ($n\in {\rm odd}$) and $J_{n}=J_{\rm e}(t)=1$  ($n\in {\rm even}$).  This time-dependent scheme is first reported to configure topological quantum networks for state transfer between distant qubits~\cite{Lang2017} . Here, we choose $N=20$ sites and set the total evolution time $T$ sufficiently large to realize the high-fidelity  transport. Figures~\ref{Fig2}(b) and ~\ref{Fig2}(c) show the gap-protected energy bands and the evolution results, respectively.

To evaluate robustness, we introduce time-independent random hopping disorders $\delta J_n \in [-\Delta,\Delta]$ in the topological structure. For each disorder strength $\Delta$, we generate 200 disorder realizations to analyze the transition amplitude.  Figures~\ref{Fig2}(d,f) and ~\ref{Fig2}(e,g) show the magnitude and the phase of the transition amplitude, respectively.
Evidently, as disorder strength increases, the average magnitude of the transition amplitude gradually decreases. Interestingly, when the disorder strength is below 0.1, the accumulated phase across 200 disorder configurations remains unchanged for each disorder strength. Importantly, through numerical analysis of transport processes with various numbers of lattice sites, we reveal the relationship between the accumulated phase and the lattice site number
\begin{equation}  
\label{SSHeven}
\gamma=
\left\{  
     \begin{array}{lr}  
     \pi/2, & N \equiv 0  \mod 4 \\
     -\pi/2, &  N \equiv 2  \mod 4
     \end{array}  
\right.  
\end{equation} 
under weak disorders.

When the disorder strength exceeds $\Delta_{c}=0.1$, the transition amplitude phase $\gamma$ randomly assumes discrete values of $-\pi/2$ or $\pi/2$, while its magnitude $|A|$ exhibits stochastic variations spanning the full interval $[0,1]$. These results indicate that the scheme is no longer viable for QST when the disorder strength exceeds the critical disorder strength $\Delta_{c}$. In numerical calculations, the critical disorder strength decreases with the number of lattice site $N$ in the transport.

\textit{Edge-defect topological transport.}---
Different from the standard SSH chain with even sites, some topological transfer protocols based on the edge-defect SSH model are proposed~\cite{Mei2018,Palaiodimopoulos2021}.
As shown in Fig.~\ref{Fig3}(a), an unpaired cell appears at the edge of the SSH model, therefore the total number of lattice sites $N$ is odd. We set $N=19$ and the hopping parameters satisfy $2J_{\rm o}(t)=1-\cos(\pi t/T)$, $2J_{\rm e}(t)=1+\cos(\pi t/T)$ and make the transfer period $T$ large enough to keep adiabaticity. Figures~\ref{Fig3}(b1,c1) display the energy bands and the excitation transmission under this kind of cosine modulations. In the presence of random disorder $\delta J_n \in [-\Delta,\Delta]$, the transfer probability can still approach unity even under extremely large disorder strength, while the accumulated phase at the receiver node remains fixed despite disorder-induced interference, as demonstrated in Fig.~\ref{Fig3}(d1-g1).

Figures~\ref{Fig3}(b2,c2) exhibit energy bands and excitation transmission of another edge-defect topological transport protocol with exponential modulation~\cite{Palaiodimopoulos2021}. In the topological scheme, $J_{\rm o}(t)=(1-e^{-\alpha t/T})/(1-e^{-\alpha})$ and $J_{\rm e}(t)=[1-e^{-\alpha (1-t/T)}]/(1-e^{-\alpha})$. We set the coefficient $\alpha=6$. By comparing Figs.~\ref{Fig3}(d1,f1) and Figs.~\ref{Fig3}(d2,f2), it is evident that the exponentially modulated protocol outperforms the cosine-modulated scheme under identical disorder strength. Importantly, both mechanisms exhibit site-number-dependent phase accumulation in the presence of disorder, as shown in Figs.~\ref{Fig3}(e1,g1,e2,g2).

By numerical calculations, we further investigate the relationship between the number of lattice sites ($N$) and the accumulated phase in this edge-defect topological model. Importantly, we find the relation
\begin{equation}  
\label{SSHodd}
\gamma=
\left\{  
     \begin{array}{lr}  
     0, & N \equiv 1  \mod 4 \\
     \pi, &  N \equiv 3  \mod 4
     \end{array}  
\right. 
\end{equation} 
persisting under strong disorder. 

Combining with Equation~(\ref{SSHeven}) , we can readily conclude that the accumulated phase at the receiver follows a $\mathbb{Z}_{4}$ symmetry group $\{e^{-i\pi/2},e^{i\pi/2},e^{i\pi},1\}$ for different site numbers in 1D topological transport protocols. As we discuss above, the site-number-dependent accumulated phase provides the precise phase correction 
\begin{equation}
\phi_0 =\frac{N-1}{2}\pi
\label{phi0}
\end{equation}
for QST in these topologically protected spin chains. Next, we check the validity of the phase correction in other 1D topological transfer schemes.

\textit{Topological interface transport.}---
Beyond edge states, another class of gap-protected QST channels is the topological interface~\cite{Yuan2021,Longhi2019}. As shown in Figs.~\ref{Fig4}(a), the intra-cell hopping at the $m$-th cell is $J_{\rm o}^{m}=\sqrt{(N_{c}-m+1)/N_{c}}$ while the inter-cell hopping is $J_{\rm e}^{m}=\sqrt{m/N_{c}}$. Therefore, there is a interface position (indicated by purple arrow) separating topologically trivial ($J_{\rm o}^{m}>J_{\rm e}^{m}$) and non-trivial regions ($J_{\rm o}^{m}<J_{\rm e}^{m}$).
Moreover, these two interaction terms are modulated by sinusoidal and cosinusoidal functions, respectively. This time-modulated inhomogeneous coupling distribution emulates the square-root operator form between Fock states~\cite{Jaynes1963,Cai2020,Yuan2021}, thereby resulting in completely flat energy bands, see Figs.~\ref{Fig4}(b). Under adiabatic conditions, the initial excitation of the sender site is perfectly transported to the receiver through adiabatic evolution guided by time-modulated topological interface states, as shown in Figs.~\ref{Fig4}(c).

To numerically study the robustness of the transition amplitude, we introduce stochastic noise to the coupling strengths. Like the foregoing process, the effective coupling strengths become $J_{\rm o,e}^{m}(t)+\delta J_{n}$, where $\delta J_n \in [-\Delta,\Delta]$ depends solely on spatial position. We set the adiabatic transfer period as $T=1000$, analyzing 200 independent disorder realizations at each disorder strength. Figures~\ref{Fig4}(d,f) demonstrate remarkable robustness of this mechanism against disorder perturbations. The phase of transition amplitude also satisfies the aforementioned phase relation (Equation~\ref{SSHodd}), as shown in Figs.~\ref{Fig4}(e,g). These results indicate that implementing this topologically protected transfer mechanism in spin chains enables robust high-fidelity QST.

Figure~\ref{Fig5}(a) illustrates another topological interface model~\cite{Longhi2019,Shen2020,Tian2022}, which includes two topologically non-trivial SSH chains with Gaussian modulations $J_{\rm o}(t)=\exp [-(t-\delta/2-T/2)^2/w^2)]$ and $J_{\rm o}^{'}(t)=\exp [-(t+\delta/2-T/2)^2/w^2)]$. Similarly, the initial excitation adiabatically follows the gap-protected channel displayed in Fig.~\ref{Fig5}(b) from the sender to the receiver. The entire evolution process is shown in Fig.~\ref{Fig5}(c). Due to the special structure, the total number of sites satisfies $N\equiv 3 \mod 4$. Figure~\ref{Fig5}(d,e) demonstrate the robustness evaluation for the transfer protocol with $N=19$. It is clear that the accumulated phase remains $\pi$ across the 200 independent disorder realizations at each disorder strength even when the transition amplitude magnitude becomes negligible.

\textit{Rice-Mele transfer schemes.}---
In addition to the SSH model, another prominent example of a topological system is the Rice-Mele lattice~\cite{Rice1982}. As shown in Fig.~\ref{Fig6}(a), there is a stagger potential energy $\lambda$ and $-\lambda$ across lattice sites. Reference~\cite{Longhi2019a} puts forward a topological transfer scheme based on the three-stage modulated Rice-Mele model. In particular, the nearest neighbor coupling
\begin{equation}  
\label{RM-Jo}
J_{\rm o}(t)=
\left\{  
     \begin{array}{lr} 
     \frac{1-\epsilon}{2}\left(1-\cos\frac{\pi t}{\tau} \right) , & 0<t<\tau \\
     1-\epsilon, & \tau<t<\tau+\tau_{z} \\
       \frac{1-\epsilon}{2}\left(1-\cos\frac{\pi t}{\tau-\tau_{z}} \right), &  \tau+\tau_{z}<t<T
     \end{array}  
\right.
\end{equation} 
and the on-site potential
\begin{equation}  
\label{RM-lambda}
\lambda(t)=
\left\{  
     \begin{array}{lr} 
     \lambda_{0}, & 0<t<\tau \\
     \lambda_{0}-\alpha (t-\tau)/2, & \tau<t<\tau+\tau_{z} \\
     -\lambda_{0}, &  \tau+\tau_{z}<t<T
     \end{array}  
\right..
\end{equation}
Here, the parameters $\tau_{z}=T-2\tau$ and $\alpha=4\lambda_{0}/\tau_{z}$. We choose $\epsilon=0.1, \tau=200, \tau_{z}=T-2\tau$, and $\lambda_{0}=0.2$ to calculate the energy bands and the transmission process in Fig.~\ref{Fig6}(b,c). It is obvious that the initial excitation at the sender site can be perfectly transferred to the receiver without noise. 

In the presence of disorder perturbations at nearest-neighbor couplings, we numerically investigate the transition amplitude of this transport mechanism. As demonstrated in Figs.~\ref{Fig6}(d,e), it is evident that under weak disorder conditions, while the magnitude of the transition amplitude retains high values, its phase exhibits stochastic behavior instead of remaining fixed. Therefore, this topologically protected transport mechanism is unsuitable for QST in spin chains, but is better suited for propagating classical excitations, particularly in scenarios where phase information is not considered.

\textit{Nonadiabatic topological transfer.}---
To satisfy the adiabaticity requirement of the topologically protected mechanism, the transfer period $T$ typically needs to be extremely long. For real quantum systems with finite lifetimes, it is challenging to implement these adiabatically modulated topological mechanisms. Recently, we proposed a non-adiabatic topological transfer without introducing additional modulation~\cite{Tian2024}. Specifically, for the aforementioned topological interface transport scheme in Fig.~\ref{Fig4}, non-adiabatic transitions between energy levels are utilized to accelerate excitation transport.

We numerically calculate the transfer probability at the receiving terminal under varying transfer periods, as shown in Fig.~\ref{Fig7}(a). Within the regime significantly below the adiabatic threshold, the transfer probability exhibits a critical period at which the transfer probability approaches unity. For the system size $N=21$, we select the critical period $T_{c}=52$ (indicated by a green triangle) and plot the transport dynamics in Fig.~\ref{Fig7}(b).
It is evident that the dynamic trajectories presented in Fig.~\ref{Fig7}(b) and Fig.~\ref{Fig4}(b) exhibit distinct characteristics. 

Furthermore, to examine the robustness of non-adiabatic topological transfer, we investigate two chains with $N=21$ and $N=19$. For each disorder strength, we compute results from 500 random disorder configurations. Comparing Fig.~\ref{Fig7}(c,e) and Fig.~\ref{Fig4}(d,f), it is apparent that non-adiabatic topological transfer is more susceptible to disorder perturbations compared to its adiabatic counterpart. However, under weak disorder conditions, non-adiabatic topological transport still shows noise-resistant behavior. More significantly, the accumulated phase of the transition amplitude in this regime adheres to the site-number-dependent relationship previously proposed in Equation~\ref{SSHodd}, as shown in Fig.~\ref{Fig7}(d,f).

Significantly, although the robustness against disorder noise in the non-adiabatic regime does not exceed that of the adiabatic regime, our calculations reveal that non-adiabatic transfer achieves a twentyfold shorter duration than adiabatic transport. These findings suggest that non-adiabatic topological transfer demonstrates superior potential for QST in spin chains.

\textit{Summary and Discussion.}---
Through revisiting QST in spin chains, we demonstrate that phase accumulation during excitation transmission constitutes a crucial factor that cannot be neglected. Furthermore, we numerically investigate various 1D topologically protected QST mechanisms, including SSH chains with cosine, Gaussian, exponential, and square-root modulation profiles. Remarkably, all these 1D topological transfer mechanisms share a universal site-dependent phase accumulation that satisfies $\mathbb{Z}_{4}$ symmetry. Under weak disorder conditions, this relationship between accumulated phase and site number persists even in non-adiabatic topological transfer protocols.
Surprisingly, this phase relationship also manifests in perfect mirror transmission mechanisms~\cite{Christandl2004,Tian2020}, which impose no constraints on spin numbers and possess a static, non-topological Hamiltonian.

Based on these findings, we uncover a universal phase correction $\phi_0 =(N-1)\pi/2$ for QST in 1D topological spin chains. Specifically, once the number of participating spins is determined, a compensatory phase gate
\begin{eqnarray}
\begin{bmatrix}
  1 & 0 \\
  0 & e^{i\pi \frac{N-1}{2}} 
\end{bmatrix}
\end{eqnarray}
can be applied at the receiving qubit to faithfully reconstruct the initial quantum state. Additionally, our work reveals that in Rice-Mele model-based topological transfer, the phase factor at the receiving qubit becomes completely randomized under noise, rendering this scheme unsuitable for high-fidelity QST applications.

\textit{Acknowledgments.}---
This work was supported by the National Natural Science Foundation of China (Grant No. 12304566) and the Fundamental Research Program of Shanxi Province (Grant No. 202303021221067). We acknowledge the use of DeepSeek for grammatical refinement during the editing phase of this paper.

\bibliographystyle{apsrev4-1}
\bibliography{QST2024}

\end{document}